\documentclass[aps,prl,preprint,showpacs,groupedaddress]{revtex4}
\usepackage {amssymb}
\usepackage {amsmath}
\usepackage {graphicx}
\usepackage {longtable}

\begin{document}
\title{Effect of strong electric fields on ferroelectric and
metal-insulator transitions}

\author{Fedor V.Prigara}
\affiliation{Institute of Physics and Technology,
Russian Academy of Sciences,\\
21 Universitetskaya, Yaroslavl 150007, Russia}
\email{fvprigara@rambler.ru}

\date{\today}

\begin{abstract}

It is shown that a sufficiently strong external electric field
causes a decrease in the transition temperature of ferroelectric,
antiferroelectric, and metal-insulator transitions. The
temperature dependence of the critical electric field suppressing
a low-temperature phase is obtained. The relation between the
low-temperature charge gap (the band gap) in the insulating phase
and the metal-insulator transition temperature is also inferred.

\end{abstract}


\maketitle

In insulators, due to the presence of localized ionic charges, a strong
external electric field produces mechanical stresses. It was shown recently
[1] that a sufficiently high mechanical stress or tension causes a decrease
in the melting temperature of a crystalline solid. Here we show that
mechanical stresses produced by a sufficiently strong electric field in a
low-temperature insulating phase lead to a decrease in the transition
temperature of ferroelectric, antiferroelectric, and metal-insulator
transitions. Such a decrease has been observed recently for the Verwey
transition in magnetite ($Fe_{3} O_{4} $) from a high-temperature ``bad
metal'' conducting phase to a low-temperature insulating phase [2]. A linear
decrease in the Curie temperature caused by a strong external electric field
is known for the antiferroelectric transition in lead zirconate ($PbZrO_{3}
$) [3]. In the case of metal-insulator transitions, the ratio of the
low-temperature charge gap (the band gap) in the insulating phase to the
transition temperature can be inferred from a general criterion of phase
transitions in crystalline solids based on a concept of the critical number
density of elementary excitations [4].

It was shown recently [1] that in a crystalline solid under a sufficiently
strong mechanical stress $\sigma = F/S$, applied to a solid, where \textit{F
}is the applied force and \textit{S} is the cross-section area of the solid
in the plane perpendicular to the direction of the applied force, the energy
$E_{v} $ of the vacancy formation decreases in accordance with the formula

\begin{equation}
\label{eq1}
E_{v} \cong E_{v0} - \alpha \sigma /n_{0} ,
\end{equation}

\noindent
where $E_{v0} $ is the energy of the vacancy formation in absence of
mechanical\textbf{} stresses,\textbf{} $n_{0} \approx 1.1 \times 10^{22}cm^{
- 3}$ and $\alpha \approx 18$ are quantum constants, and $\sigma $ is the
absolute value of the mechanical stress. A thermodynamic consideration based
on the Clausius- Clapeyron equation gives the number density $n_{v}
$\textit{} of vacancies in a solid in the form [1]

\begin{equation}
\label{eq2}
n_{v} = \left( {P_{0} /T} \right)exp\left( { - E_{v} /T} \right) = \left(
{n_{0} T_{0} /T} \right)exp\left( { - E_{v} /T} \right),
\end{equation}

\noindent
where $E_{v} $ is the energy of the vacancy formation, $P_{0} = n_{0} T_{0}
$ is a constant, $T_{0} $ can be put equal to the melting temperature of the
solid at ambient pressure (in absence of mechanical stresses), and the
Boltzmann constant $k_{B} $ is included in the definition of the temperature
\textit{T}.

The melting of a crystalline solid occurs when the critical number density
$n_{c} $ of vacancies is achieved, the critical number density of vacancies
being [4]

\begin{equation}
\label{eq3}
n_{c} \cong n_{0} exp\left( { - \alpha}  \right).
\end{equation}

The equations (\ref{eq2}) and (\ref{eq3}) give the relation between the energy $E_{v} $ of
the vacancy formation and the melting temperature $T_{m} $ of a crystalline
solid in the form

\begin{equation}
\label{eq4}
E_{v} \cong \alpha T_{m} .
\end{equation}

A decrease in the energy of the vacancy formation and, whence, in the
melting temperature is produced by the strong atomic relaxation in a
crystalline solid under a sufficiently strong mechanical stress.

This concept can be extended on other phase transitions, such as
ferroelectric, ferromagnetic, and superconducting, if we consider the
corresponding phase transition as the melting of some lattice [4]. In the
case of the ferroelectric transition, it is the lattice of electrical
dipoles; in the case of the ferromagnetic transition, it is the lattice of
magnetic dipoles. Consider now the number density \textit{n} of elementary
excitations in a corresponding lattice, instead of the number density of
vacancies in an ordinary lattice. An elementary excitation invokes the
change of the sign of the electrical or magnetic dipole moment of the
elementary lattice cell for ferroelectrics and ferromagnets, respectively.

Similarly to the energy of the vacancy formation (\ref{eq1}), the energy \textit{E}
of an elementary excitation depends on the absolute value $\sigma $of the
mechanical stress (in the region of strong mechanical stresses) as follows

\begin{equation}
\label{eq5}
E = E_{0} - \alpha _{P} \sigma /n_{0} ,
\end{equation}

\noindent
where $E_{0} $ is the energy of an elementary excitation at ambient pressure
(in absence of mechanical stresses), and $\alpha _{P} $ is a dimensionless
constant dependent on the type of the phase transition [4]. The atomic
relaxation constant $\alpha _{P} $ is related to a dimensionless pressure
derivative of the transition temperature $T_{c} $ (in the region of high
pressures) as follows

\begin{equation}
\label{eq6}
k = n_{0} dT_{c} /dP \cong - \alpha _{P} /\alpha .
\end{equation}

Here \textit{P} is the hydrostatic pressure and the values of \textit{k} and
$\alpha _{P} $ for ferroelectrics are $k \approx - 1/9$ and $\alpha _{P}
\approx - k\alpha \approx 2$[4].

It can be inferred from the experimentally measured pressure derivative of
the metal-insulator transition temperature in $BaVS_{3} $ [5] that $k
\approx - 1/9$ and $\alpha _{P} \approx - k\alpha \approx 2$ for
metal-insulator transitions too. Note that both ferroelectric and
metal-insulator transitions are associated with a lattice distortion of a
ferroelastic type (see, e.g., [6]).

The number density \textit{n} of elementary excitations is given by the
formula analogous to the equation (\ref{eq2}),

\begin{equation}
\label{eq7}
n = \left( {n_{0} T_{c} /T} \right)exp\left( { - E/T} \right),
\end{equation}

\noindent
where $T_{c} $ denotes here the transition temperature at ambient pressure
(in absence of mechanical stresses and external fields).

The corresponding phase transition occurs when the critical number density
$n_{c} $ of elementary excitations (given by the equation (\ref{eq3})) is achieved
[4], so that the energy \textit{E} of an elementary excitation is related to
the transition temperature $T_{c} $, similarly to the equation (\ref{eq4}),

\begin{equation}
\label{eq8}
E \cong \alpha T_{c} .
\end{equation}

In the case of metal-insulator transitions, the energy \textit{E} of an
elementary excitation is equal to the charge gap $\Delta _{ch} $ (the band
gap $E_{g} $). Optical reflectivity measurements in $BaVS_{3} $ [5] show
that the ratio of the charge gap $\Delta _{ch} = E_{g} = E$ to the
metal-insulator transition temperature $T_{MI} $ is constant along the phase
equilibrium curve in the temperature-pressure plane, in agreement with the
relation (\ref{eq8}). Since the charge gap is temperature dependent, it is a
low-temperature value $\Delta _{ch} \left( {0} \right)$of the charge gap
which should be compared with the value of the charge gap given by the
equation (\ref{eq8}). In $BaVS_{3} $ at ambient pressure, the ratio of the charge
gap to the metal-insulator transition temperature is $\Delta _{ch} \left(
{0} \right)/T_{MI} \cong 17$ at 10 K and $\Delta _{ch} /T_{MI} \cong 12$ at
60 K, as it can be determined from the low-frequency optical conductivity
data reported in Ref.5. (The transition temperature $T_{MI} $ at ambient
pressure is $T_{MI} \approx 70K$).

In germanium \textit{Ge}, the band gap is equal $E_{g} \left( {0} \right) =
0.744eV$, and the relation (\ref{eq8}) gives the metal-insulator transition
temperature $T_{MI} \cong 500K$.

Multiferroic bismuth ferrite ($BiFeO_{3} $) exhibits a
metal-insulator transition from the metallic cubic $\gamma$ phase
to the insulating ferroelastic orthorhombic $\beta $ phase at
$T_{MI} \approx 1200K \approx 0.103eV$ [7]. The relation
(\ref{eq8}) gives a band gap $E_{g} \cong \alpha T_{MI} \cong
1.85eV$. Optical absorption measurements show that the band gap of
bismuth ferrite decreases slowly and linearly with temperature in
the $\alpha $ and $\beta $ phases from $2.5eV$to $1.5eV$ [7]. (The
rhombohedral $\alpha $ phase which is stable below 1100 \textit{K}
is ferroelectric and undergoes an antiferromagnetic transition at
$T_{N} \approx 640K$).

In cadmium chalcogenides, the band gap $E_{g} $ is proportional to the
melting temperature $T_{m} $ or, equivalently, to the energy $E_{v} $ of the
vacancy formation, $E_{g} \cong \left( {3/4} \right)E_{v} $:

\newcommand{\PreserveBackslash}[1]{\let\temp=\\#1\let\\=\temp}
\let\PBS=\PreserveBackslash
\begin{longtable}
{|p{71pt}|p{71pt}|p{71pt}|p{71pt}|p{71pt}|p{71pt}|} a & a & a & a
& a & a  \kill \hline
Material &
 $T_{m} ,$ \textit{K}&
 $E_{v} ,$ \textit{eV}&
 $E_{g} ,$ \textit{eV}&
\[
E_{g} /E_{v}
\]
&
Crystal structure \\
\hline
\textbf{}CdS&
2020&
3.13&
2.42&
0.77&
hexagonal \\
\hline
\textbf{}CdSe&
1541&
2.40&
1.74&
0.73&
hexagonal \\
\hline
\textbf{}CdTe&
1365&
2.12&
1.56&
0.74&
cubic \\
\hline
\end{longtable}

\textbf{}

Similar is the ratio $E_{g} /E_{v} $ in diamond (0.8) and $\alpha - Al_{2}
O_{3} $ (0.7). In silicon (Si) and germanium (Ge), the band gap is $E_{g}
\cong 0.4E_{v} $.

We assume that a sufficiently strong electric field ${\rm E}$ applied to an
insulating material produces a mechanical stress $\sigma $\textbf{} which
depends linearly on the electric field\textbf{} ${\rm E}$ as given by the
formula

\begin{equation}
\label{eq9}
\sigma = \beta {\rm E}.
\end{equation}

The coefficient $\beta $ will be estimated below. Substituting expression
(\ref{eq9}) in the equation (\ref{eq5}) for the energy \textit{E} of an elementary
excitation we obtain

\begin{equation}
\label{eq10}
E = E_{0} - \alpha _{P} \beta {\rm E}/n_{0} .
\end{equation}

Now the equations (\ref{eq7}) and (\ref{eq3}) give the phase equilibrium curve in the
temperature-electric field plane in the form

\begin{equation}
\label{eq11}
\left( {E_{0} - \alpha _{P} \beta {\rm E}/n_{0}}  \right)/T \cong \alpha
\cong E_{0} /T_{c} .
\end{equation}

The last equation gives the temperature dependence of the critical electric
field ${\rm E}_{c} $ suppressing a low-temperature phase as follows

\begin{equation}
\label{eq12}
{\rm E}_{c} \cong {\rm E}_{c} \left( {0} \right)\left( {1 - T/T_{c}}
\right),
\end{equation}

\noindent
where the value of the critical electric field at zero temperature is

\begin{equation}
\label{eq13}
{\rm E}_{c} \left( {0} \right) \cong \left( {n_{0} E_{0}}  \right)/\left(
{\alpha _{P} \beta}  \right).
\end{equation}

A mechanical stress $\sigma $ produced by a sufficiently strong electric
field\textbf{} ${\rm E}$ in an insulating material with ionic charges $z_{i}
e$ in the unit cell (\textit{e} is the charge of an electron and \textit{i}
is the number of an ion) can be estimated as follows

\begin{equation}
\label{eq14}
\sigma \cong Ze{\rm E}/a_{0}^{2} ,
\end{equation}

\noindent
where \textit{Z} is the total sum of absolute values of ionic charges in the
unit cell (in units of the electron charge \textit{e}), $Z =
\sum\nolimits_{i} {\left| {z_{i}}  \right|} $, and $a_{0} = n_{0}^{ - 1/3}
\cong 0.45nm$ has an order of the lattice parameter \textit{a}. Here it is
assumed that an external electric field acts only on a charged layer at the
surface of an insulating material (in the bulk the field is much weaker).

In lead zirconate ($PbZrO_{3} $), a linear temperature dependence of the
critical electric field for the antiferroelectric transition with the Curie
temperature $T_{c} \approx 500K$ is indeed observed [3], the extrapolated
value of the critical electric field at zero temperature being ${\rm E}_{c}
\left( {0} \right) \cong 800kV/cm$. In this case $Z = 12$per formula.

The equation (\ref{eq14}) gives an estimation of the coefficient $\beta $ in the
form

\begin{equation}
\label{eq15}
\beta \cong Ze/a_{0}^{2} .
\end{equation}

Substituting this expression in the equation (\ref{eq13}), we obtain

\begin{equation}
\label{eq16}
{\rm E}_{c} \left( {0} \right) \cong \left( {\alpha /\alpha _{P} Z}
\right)\left( {T_{c} /ea_{0}}  \right).
\end{equation}

In the case of lead zirconate ($PbZrO_{3} $)$T_{c} \approx
500K$,$Z = 12$, and the equation (\ref{eq16}) gives the value
${\rm E}_{c} \left( {0} \right) \cong 750kV/cm$ which is close to
the extrapolated experimental value of the zero-temperature
critical electric field indicated above.

An estimate (\ref{eq14}) is valid only for sufficiently strong electric fields. In
the case of lead zirconate, ${\rm E} \geqslant 20kV/cm$; for weak electric
fields, ${\rm E} < 20kV/cm$, the Curie temperature $T_{c} $ is independent
of the electric field ${\rm E}$. In general case, the mechanical response in
weak electric fields depends on the crystal symmetry [3]. The relation (\ref{eq14})
cannot apply to the metallic conducting phase above the metal-insulator
transition.

Polarization in thin films of $BiFeO_{3} $ grown on $SrTiO_{3} $ substrates
[7] has an order of

\begin{equation}
\label{eq17}
P_{r} \cong \beta /\left( {4\pi}  \right) \cong \left( {Z/4\pi}
\right)\left( {e/a_{0}^{2}}  \right) \cong e/a_{0}^{2} \cong 80\mu
C/cm^{2},
\end{equation}

\noindent where $Z = 12$for $BiFeO_{3} $. The critical electric
field given by the equation (\ref{eq16}) is much weaker than the
effective polarization field (displacement) $4\pi P_{r} $. This
rules out a direct effect of an external electric field on the
energy of an elementary excitation.

In $PbZrO_{3} $, an external electric field ${\rm E} \geqslant 20kV/cm$
produces a transition from the paraelectric cubic phase to the ferroelectric
rhombohedral phase above the antiferroelectric Curie temperature $T_{c}
\approx 500K$ and a transition from the antiferroelectric orthorhombic phase
to the ferroelectric rhombohedral phase below the antiferroelectric Curie
temperature. Sufficiently strong electric fields suppress the ferroelectric
phase as it follows from the above consideration. This conclusion is
supported by the phase diagram of $Pb\left( {Zr,Ti} \right)O_{3} $ system
[3]. The Ti doping in $PbZrO_{3} $ acts as an external electric field
stabilizing the rhombohedral ferroelectric phase. However, a sufficiently
high Ti doping suppresses this rhombohedral phase.

In $BaTiO_{3} $, a weak external electric field (${\rm E} < 1kV/cm$)
produces an increase in the ferroelectric Curie temperature [3], but
stronger electric fields suppress a ferroelectric phase, in accordance with
the above consideration.

The critical electric field ${\rm E}_{c} \left( {0} \right) \cong 800kV/cm$
in $PbZrO_{3} $ has an order of the breakdown field in thin $BiFeO_{3} $
films [8]. The breakdown field can be estimated as follows. According to the
equations (\ref{eq1}) and (\ref{eq9}), a sufficiently strong external electric field ${\rm
E}$ produces a decrease in the energy $E_{v} $ of the vacancy formation and,
whence, the decrease in the melting temperature $T_{m} $ of an insulating
material. The melting curve of an insulator in strong electric fields is
determined by the equation similar to the equation (\ref{eq11}),

\begin{equation}
\label{eq18}
\left( {E_{v0} - \alpha \beta {\rm E}/n_{0}}  \right)/T \cong \alpha \cong
E_{v0} /T_{m} .
\end{equation}

The last equation gives the temperature dependence of the melting field (or
the decomposition field) ${\rm E}_{m} $ in the form

\begin{equation}
\label{eq19}
{\rm E}_{m} \cong {\rm E}_{m} \left( {0} \right)\left( {1 - T/T_{m}}
\right),
\end{equation}

\noindent
where the zero-temperature value of the melting field is

\begin{equation}
\label{eq20}
{\rm E}_{m} \left( {0} \right) \cong n_{0} E_{v0} /\alpha \beta \cong n_{0}
T_{m} /\beta .
\end{equation}

Substituting an estimate (\ref{eq15}) for the coefficient $\beta $ in the equation
(\ref{eq20}), we obtain

\begin{equation}
\label{eq21}
{\rm E}_{m} \left( {0} \right) \cong \left( {1/Z} \right)\left( {T_{m}
/ea_{0}}  \right).
\end{equation}

In the case of bismuth ferrite $BiFeO_{3} $, the decomposition
temperature is $T_{d} \approx 0.11eV$ [7] and $Z = 12$ per
formula, so that the equation (\ref{eq21}) gives ${\rm E}_{d}
\left( {0} \right) \cong 200kV/cm$. The breakdown field in thin
$BiFeO_{3} $ films is about ${\rm E}_{b} \cong 700kV/cm$ [8]. The
electrical breakdown in $BiFeO_{3} $ thin films is associated with
a phase separation of $BiFeO_{3} $ into magnetite $Fe_{3} O_{4} $.
The coercive field in $BiFeO_{3} $ thin films is also very high,
about three times larger than in other ferroelectric thin films
[8]. The breakdown field for Sm-doped bismuth titanate thin films
is about ${\rm E}_{b} \cong 500kV/cm$ [9].

To summerize, we obtained the dependence of the energy of an elementary
excitation on the electric field for ferroelectrics, antiferroelectrics, and
paraelectric insulators in the region of strong electric fields. We obtained
the temperature dependence of the critical electric field suppressing a
low-temperature phase for ferroelectric, antiferroelectric, and
metal-insulator transitions. An estimate of the breakdown field in
insulating materials is given. The relation between the charge gap (the band
gap) in the insulating phase and the metal-insulator transition temperature
is also obtained.

\begin{center}
---------------------------------------------------------------
\end{center}

[1] F.V.Prigara, E-print archives, cond-mat/0701148 (2007).

[2] S.Lee, A.Fursina, J.T.Mayo, C.T.Yavuz, V.L.Colvin, R.G.S.Sofin,
I.V.Shvets, and D.Natelson, Nature Materials \textbf{7}, 130 (2008).

[3] G.S.Zhdanov, \textit{Solid State Physics} (Moscow University Press,
Moscow, 1961).

[4] F.V.Prigara, E-print archives, arXiv: 0708.1230 (cond-mat.supr-con)
(2007).

[5] I.Kezsmarki, G.Mihaly, R.Gaal, N.Barisic, H.Berger, L.Forro, C.C.Homes,
and L.Mihaly, Phys. Rev. B \textbf{71}, 193103 (2005).

[6] S.H.Curnoe and A.E.Jacobs, Phys. Rev. B \textbf{64}, 064101 (2001).

[7] R.Palai, R.S.Katiyar, H.Scmid et al., Phys. Rev. B
\textbf{77}, 014110 (2008).

[8] X.J.Lou, C.X.Yang, M.Zhang, and J.F.Scott, Appl. Phys. Lett.
\textbf{90}, 262908 (2007).

[9] X.J.Lou, X.Hu, M.Zhang, S.A.T.Redfern, and J.F.Scott, Integrated
Ferroelectrics \textbf{73}, 93 (2005).

\end{document}